\documentclass[showpacs,prb,aps,twocolumn]{revtex4}
\usepackage{graphicx}
\usepackage[normalem]{ulem}
\usepackage{amssymb}
\usepackage{dcolumn}% Align table columns on decimal point

\hfuzz=\maxdimen
\begin{document}

\title{Effect of iron content and potassium substitution in
A$_{0.8}$Fe$_{1.6}$Se$_2$ (A = K, Rb, Tl) superconductors: a
Raman-scattering investigation}
\author{A. M. Zhang, K. Liu, J. B. He, D. M. Wang, G. F. Chen, B. Normand}
\author{Q. M. Zhang}
\email{qmzhang@ruc.edu.cn}
\affiliation{Department of Physics, Renmin University of China, Beijing
100872, P. R. China}

\date{\today}

\begin{abstract}

We have performed Raman-scattering measurements on high-quality single
crystals of the superconductors K$_{0.8}$Fe$_{1.6}$Se$_2$ ($T_c$ = 32 K),
Tl$_{0.5}$K$_{0.3}$Fe$_{1.6}$Se$_2$ ($T_c$ = 29 K), and
Tl$_{0.5}$Rb$_{0.3}$Fe$_{1.6}$Se$_2$ ($T_c$ = 31 K), as well as of the
insulating compound KFe$_{1.5}$Se$_2$. To interpret our results, we have
made first-principles calculations for the phonon modes in the ordered
iron-vacancy structure of K$_{0.8}$Fe$_{1.6}$Se$_2$. The modes we observe
can be assigned very well from our symmetry analysis and calculations,
allowing us to compare Raman-active phonons in the AFeSe compounds. We
find a clear frequency difference in most phonon modes between the
superconducting and non-superconducting potassium crystals, indicating
the fundamental influence of iron content. By contrast, substitution of
K by Tl or Rb in A$_{0.8}$Fe$_{1.6}$Se$_2$ causes no substantial frequency
shift for any modes above 60 cm$^{-1}$, demonstrating that the alkali-type
metal has little effect on the microstructure of the FeSe layer. Several
additional modes appear below 60 cm$^{-1}$ in Tl- and Rb-substituted
samples, which are vibrations of heavier Tl and Rb ions. Finally, our
calculations reveal the presence of ``chiral'' phonon modes, whose
origin lies in the chiral nature of the K$_{0.8}$Fe$_{1.6}$Se$_2$ structure.

\end{abstract}


\maketitle

\section{Introduction}

Iron pnictide superconductors display the highest superconducting transition
temperatures yet known outside cuprate systems. Unsurprisingly, their
discovery almost four years ago ignited an enduring drive both to search
for new superconducting materials and to explore the fundamental physical
properties of these systems, especially the pairing mechanism. Until
recently, five such systems had been synthesized and studied, namely
LnFeAsOF (known as ``1111,'' with Ln $\equiv$ La, Ce, Pr, Nd, Sm,
\dots),\cite{1111} AEFe$_2$As$_2$ and AFe$_2$As$_2$ (``122,'' with AE an
alkaline earth and A an alkali metal),\cite{122} AFeAs (``111''),\cite{111}
Fe(Se,Te) (``11''),\cite{11} and Sr$_2$VO$_3$FeAs (``21311'').\cite{21311}

FeSe is of particular interest among these systems for a number of reasons.
The most important is that it does not contain the poisonous element As. In
addition, its transition temperature, $T_c$, displays a very strong pressure
dependence. At ambient pressure, $T_c \approx$ 8 K,\cite{MKWu} much lower than
in the 1111 and 122 systems, but a maximum $T_c$ of 37 K can be reached under
a pressure of approximately 6 GPa.\cite{FeSe37K} It has been shown\cite{MKWu}
that the microscopic effect of the applied pressure is to alter the separation
of the Se atoms from the Fe planes, and this very strong dependence opens
the possibility of raising $T_c$ by the introduction of internal chemical
pressure. The first successful execution of this program was reported in
Ref.~\onlinecite{XLChen}, where a potassium-intercalated FeSe superconductor
was synthesized and found to have $T_c \approx 31$ K, a value comparable to
that in the 122 materials.

In parallel with intensive efforts to synthesize further examples of
A$_x$Fe$_{2-y}$Se$_2$ systems, the electronic and magnetic properties of
these compounds have been studied extensively. Infrared optical conductivity
measurements indicated that the non-superconducting system is a small-gap
semiconductor rather than a Mott insulator.\cite{Infrared} In superconducting
samples, nuclear magnetic resonance (NMR) measurements found very narrow line
widths, singlet superconductivity with no coherence peak, and only weak spin
fluctuations, but no sign of magnetism.\cite{NMR} Angle-resolved photoemission
spectroscopy (ARPES) measurements were initially inconclusive, but
now\cite{newARPES} indicate three electron-like Fermi surfaces (two
around the $\Gamma$ point and one around the M point) with full gaps on
at least two, but again no evidence for magnetic order. Both Raman\cite{ZAM}
and infrared\cite{Infrared} spectroscopy find large numbers of phonon modes
beyond those expected in a 122 structure, and transmission electron microscopy
(TEM)\cite{rTEM} reveals a well-defined surface vacancy ordering.

The first piece of the puzzle concerning the true nature of the
A$_x$Fe$_{2-y}$Se$_2$ materials was revealed by neutron diffraction
experiments.\cite{wbnd} First, these determine that the predominant
structure is dictated by a real Fe content of 1.6 in the superconducting
crystals. This gives a regular, 1/5-depleted Fe vacancy ordering pattern
with a $\sqrt{5}\times\sqrt{5}$ unit cell. Second, the magnetic properties
of this phase are perhaps the most unusual of any known superconductor,
featuring an ordered spin structure of antiferromagnetically coupled
four-spin blocks, a very high N\'eel temperature of 520 K, and an
extraordinarily large local moment of 3.31$\mu_B$ per Fe
site.\cite{wbnd} The magnetic transition has now been confirmed by bulk
measurements,\cite{XHChen} while M\"ossbauer spectroscopy has been used
to verify the large local moment, giving results of 2.9 and 2.2$\mu_B$ in
two separate studies.\cite{Mossbauer} These observations demonstrate
directly that the AFeSe superconductors are completely different from
FeAs-based and cuprate superconductors in at least two respects. One is
that a bulk ordering of the Fe vacancies plays a key role in determining
the electronic and magnetic properties of the system. The other is the
apparent (micro- or mesoscopic) coexistence of long-range antiferromagnetic
order with superconductivity.

The question of coexistence is the other piece of the puzzle. It involves
reconciling the NMR and ARPES results, which appear to originate from a
homogeneous, nonmagnetic bulk superconductor, with the data from all of
the other techniques cited above, which are the signatures of a magnetic
insulator with a complex structure. The nature of this coexistence or
cohabitation has been the focus of almost all recent experimental
investigations of the A$_x$Fe$_{2-y}$Se$_2$ materials. A clear consensus
has emerged in support of a phase separation between antiferromagnetic and
superconducting regions, but occurring on microscopic length scales. Phase
separation has been reported in optical\cite{Charnukha} and ARPES
experiments,\cite{Chen} the former authors attributing a much larger direct
band gap (0.45 eV) to the majority insulating phase than that deduced in
Ref.~\onlinecite{Infrared}. The appearance of phase separation at nanometer
scales has been detected by M\"ossbauer,\cite{Ksenofontov} X-ray,\cite{Ricci}
and in-plane optical spectroscopy measurements.\cite{Yuan} Scanning Tunneling
Microscopy (STM) has been used to image this nanoscopic phase separation
directly in epitaxially grown films.\cite{Li} Estimates of the volume
fraction of the magnetic and insulating phase by these techniques remain
close to the value of 90\% reported by muon spin resonance ($\mu$SR)
measurements.\cite{MuSR} The minority (10\%) superconducting phase must
clearly be percolating to give the appearance of bulk superconductivity.
Several authors have suggested\cite{Chen,Li,Friemel} that the ordered
vacancy configuration is present only in the AF phase, consistent with
the indications that the unusual magnetism may be an essential component
in stabilizing this structure, while the superconducting phase is structurally
homogeneous and may be composed of stoichiometric AFe$_2$Se$_2$. Finally,
there are only two experimental reports concerning the question of whether
this coexistence is collaborative or competitive; our own data from
two-magnon Raman scattering\cite{ZAM2} and additional results from neutron
diffraction\cite{wbndu} suggest a strong competition, in that 5$-$10\%
of the magnetic volume is suppressed by an apparent proximity effect at
the onset of superconductivity.

Returning to the question of sample synthesis, rapid progress followed
the first report of K$_x$Fe$_{2-y}$Se$_2$, with several groups achieving
superconductivity by substitution of alkali-type metals including Rb, Cs,
and Tl.\cite{AFeSe} The purpose of substitution by ions of equal valence
but different radii is to alter the chemical pressure to control the
electronic properties. An example is the maximum $T_c$ of approximately
56 K achieved by the substitution of rare-earth ions in the 1111
system.\cite{1111Max} It is thus somewhat surprising that substitution
of K by Rb, Cs, or Tl in the new superconductors leaves $T_c$ essentially
unaltered at around 30 K.\cite{AFeSe} This result poses another fundamental
question, concerning why superconductivity should be so robust in the AFeSe
system, and its answer requires a careful investigation into the effect of
Tl, Rb, and Cs substitution on the microstructure of the FeSe layers in
these materials.

In this paper, we address these questions through a Raman-scattering study.
We have measured the spectra in three high-quality superconducting crystals
of Tl- and Rb-substituted K$_{0.8}$Fe$_{1.6}$Se$_2$, and in one
non-superconducting crystal with an altered Fe content. For each crystal,
we observe double-digit numbers of phonon modes, dramatically different
from a normal 122 structure but consistent with an ordered vacancy
structure. We perform first-principles calculations for the zone-center
phonons in K$_{0.8}$Fe$_{1.6}$Se$_2$ ($T_c$ = 32 K), in order to assign
the observed modes by the symmetries and frequencies we measure. The
resulting assignment is very satisfactory, demonstrating that this
vacancy-ordered structure is indeed the majority phase of our samples.
From this understanding, we find the effect of a varying Fe content to be
detectable as frequency shifts of the Raman modes above 60 cm$^{-1}$, as
these are vibrations involving Fe and Se atoms. By contrast, the effects
of Tl and Rb substitution are not discernible above 60 cm$^{-1}$, indicating
that K-layer substitution causes no substantial distortion of the FeSe layer.
Below 60 cm$^{-1}$, additional modes associated with vibrations of the
heavier Tl and Rb ions can be observed in the substituted samples. In our
calculations we also find some unconventional ``chiral'' phonon modes, which
arise due to the chiral nature of the $\sqrt{5}\times\sqrt{5}$ Fe-vacancy
structure, and we consider their implications for coupling to possible chiral
electronic and magnetic modes.

The structure of the manuscript is as follows. In Sec.~II we discuss our
sample preparation and measurement techniques. In Sec.~III we present the
full theoretical analysis for computing the phonon spectrum from the known
lattice structure of the insulating and magnetic majority phase, and we
discuss the nature of the predicted modes. With this frame of reference,
we may then understand our Raman-scattering results, which are presented
in detail in Sec.~IV. Section V contains a short summary and conclusion.

\section{Materials and Methods}

The FeSe-based crystals used in our measurements were grown by the Bridgman
method. The detailed growth procedure may be found elsewhere.\cite{GFChen}
The accurate determination of crystal stoichiometry has been found to be a
delicate issue, which is crucial in establishing the proper starting point
for understanding both the magnetism and the superconductivity. We have
obtained highly accurate results for our crystal compositions by using
inductively coupled plasma atomic emission spectroscopy (ICP-AES), and
have obtained results completely consistent with the neutron diffraction
refinement.\cite{wbnd} The crystals we used in this study were
K$_{0.8}$Fe$_{1.6}$Se$_2$ ($T_c \approx$ 32 K), Tl$_{0.5}$K$_{0.3}$Fe$_{1.6}$Se$_2$
($T_c \approx$ 29 K), and Tl$_{0.5}$Rb$_{0.3}$Fe$_{1.6}$Se$_2$ ($T_c \approx$ 31
K), all of which were superconducting with similar transition temperatures,
and also the non-superconducting compound KFe$_{1.5}$Se$_2$. The precise
chemical formula for this series of compounds is thought to be
A$_x$Fe$_{2-x/2}$Se$_2$ (A = K, Rb, Cs, Tl),\cite{wbnd,rblhchgqwl} and our
results agree with this deduction. The above discussion of phase separation
notwithstanding, X-ray diffraction patterns obtained for our crystals show
no discernible secondary phases, indicating that the volume fraction of the
superconducting minority phase is low in all cases. The resistivities of the
samples were measured with a Quantum Design physical properties measurement
system (PPMS), and the magnetization by using the PPMS vibrating sample
magnetometer (VSM). The sharp superconducting and diamagnetic transitions,
which were found for all three superconducting crystals, are presented in
Sec.~IV to accompany a more detailed discussion of phase separation. These
results indicate that all of the crystals used in our Raman measurements are
of very high quality, which in a phase-separation context means that the
nanoscale percolation of the minority phase is good and homogeneous.

\begin{figure}[t]
\includegraphics[angle=0,scale=0.56]{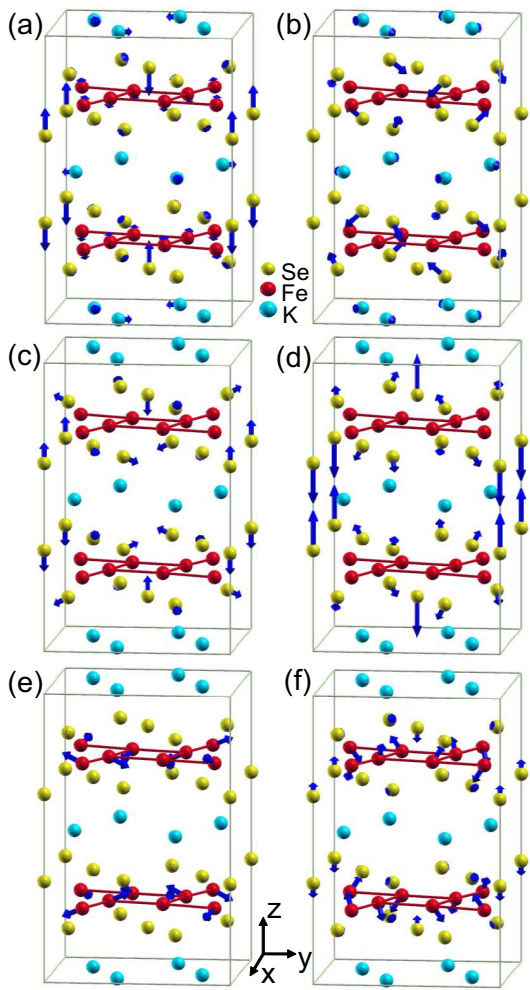}
\caption{(Color online) Atomic displacement patterns for selected Raman-active
$A_{g}$ modes of K$_{0.8}$Fe$_{1.6}$Se$_{2}$, with frequencies of (a) 75.1, (b)
130.5, (c) 159.2, (d) 212.6, (e) 268.5, and (f) 286.1 cm$^{-1}$. Fe atoms
connected by red lines have right-handed chirality in this representation.}
\label{fig1}
\end{figure}

\begin{figure}[t]
\includegraphics[angle=0,scale=0.57]{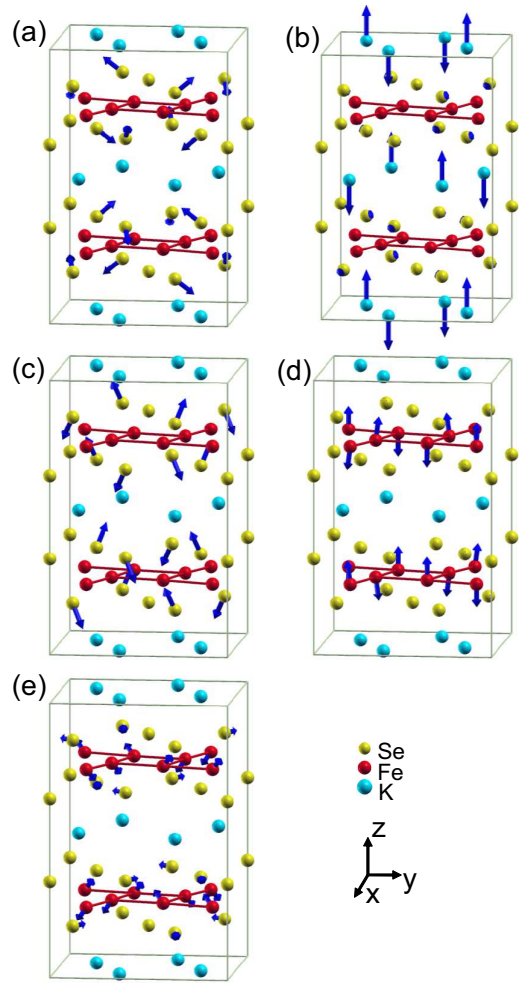}
\caption{(Color online) Atomic displacement patterns for selected Raman-active
$B_{g}$ modes of K$_{0.8}$Fe$_{1.6}$Se$_{2}$ with frequencies of (a) 66.7, (b)
106.2, (c) 149.0, (d) 238.3, and (e) 279.0 cm$^{-1}$. Fe atoms connected by
red lines have right-handed chirality.} \label{fig2}
\end{figure}

\begin{figure}[t]
\includegraphics[angle=0,scale=0.53]{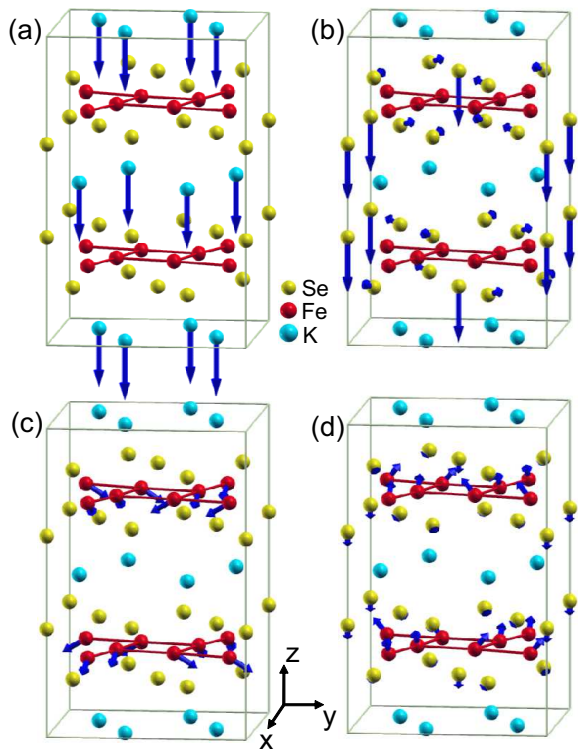}
\caption{(Color online) Atomic displacement patterns for selected
infrared-active phonon modes of K$_{0.8}$Fe$_{1.6}$Se$_{2}$ with frequencies
of (a) 119.1, (b) 212.3, (c) 253.4, and (d) 308.5 cm$^{-1}$. Fe atoms
connected by red lines have right-handed chirality.} \label{fig3}
\end{figure}

\begin{figure}[t]
\includegraphics[angle=0,scale=0.38]{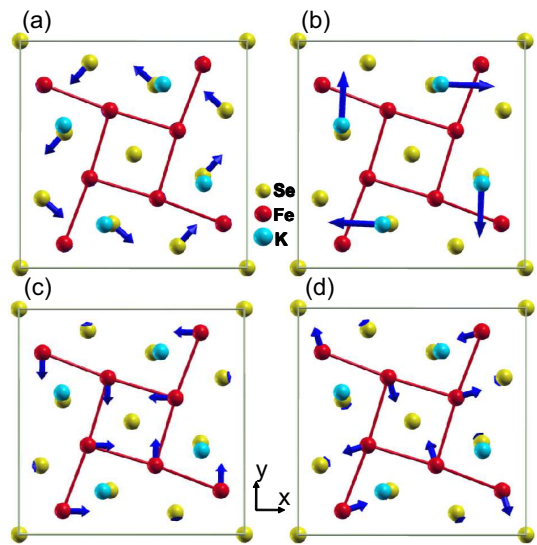}
\caption{(Color online) Atomic displacement patterns for chiral phonon modes
[(a) 67.0, (b) 86.2, and (c) 301.3 cm$^{-1}$] and for the breathing mode [(d)
269.6 cm$^{-1}$] of K$_{0.8}$Fe$_{1.6}$Se$_{2}$. Fe atoms connected by red lines
have right-handed chirality.}
\label{fig4}
\end{figure}

All measurements were made by first cleaving the crystals in a glove box, to
obtain flat, shiny $(ab)$-plane surfaces. The freshly-cleaved crystals were
sealed under an argon atmosphere and transferred into the cryostat within 30
seconds for immediate evacuation to a work vacuum of approximately $10^{-8}$
mbar. Raman-scattering measurements were performed with a triple-grating
monochromator (Jobin Yvon T64000) in a pseudo-backscattering configuration.
The beam of the 532 nm solid-state laser (Torus 532, Laser Quantum) was
focused into a spot on the sample surface with a diameter of approximately
20 $\mu$m. The beam power was reduced to avoid heating, and was kept below
0.6 mW during our measurements at the lowest temperatures; the real
temperature in the spot was deduced from the intensity relation between
the Stokes and anti-Stokes spectra. The polarization determination for the
phonons whose symmetries we assign as $A_g$ and $B_g$ in the spectra shown
in Sec.~IV was performed by adjusting the polarization of the incident and
scattered light, rather than by a formal symmetry analysis, as discussed
in detail in Ref.~\onlinecite{ZAM}.

\section{First-principles dynamical analysis}

A full understanding of our Raman-scattering results, and in particular of
the effects caused by iron content and potassium substitution, requires a
complete phonon mode assignment. The ordered pattern of Fe vacancies\cite{wbnd}
explains quite naturally the large number of optical phonons observed in
light-scattering experiments. However, the large unit cell means that a
detailed vibration analysis is somewhat involved. We begin with the results
from neutron diffraction,\cite{wbnd} which gives the structural space group
of K$_{0.8}$Fe$_{1.6}$Se$_2$ as $I4/m$ and the Wyckoff positions of the atoms
as 8h for potassium, 16i for iron, 4d for the iron vacancies, and 16i for
selenium. The corresponding symmetry analysis allows a total of 17 $A_g$ or
$B_g$ modes.\cite{ZAM}

We have calculated the nonmagnetic electronic structure and the zone-center
phonons of K$_{0.8}$Fe$_{1.6}$Se$_{2}$ from first principles by performing
density-functional calculations. We use the Vienna {\it ab-initio} simulation
package,\cite{kresse,paw} which makes use of the projector augmented wave
(PAW) method\cite{paw} combined with a general gradient approximation (GGA),
implemented with the Perdew-Burke-Ernzerhof formula,\cite{pbe} for the
exchange-correlation potentials. The nonmagnetic K$_{0.8}$Fe$_{1.6}$Se$_{2}$
system was modeled by adopting a parallelepiped supercell containing 8 Fe
atoms plus 2 Fe vacancies, 10 Se atoms, and 4 K atoms plus 1 K vacancy. The
Brillouin zone of the supercell was sampled with an $8\times8\times8$
{\bf k}-space mesh and the broadening was taken to be Gaussian. The energy
cutoff for the plane waves was 400 eV. Both the shape and volume of the cell
and the internal coordinates of all the ions were fully optimized until the
forces on all relaxed atoms were below 0.01 eV/\AA.

The frequencies and displacement patterns of the phonon modes were calculated
using the dynamical matrix method,\cite{liu05} in which the derivatives were
taken from the finite differences in atomic forces at a fixed atomic
displacement of 0.01~\AA. All 22 atoms in the supercell were allowed to
move from their equilibrium positions in all directions ($x$, $y$, $z$),
leading to a $66\times66$ matrix. The phonon frequencies and displacement
patterns are given by diagonalizing this matrix. Convergence tests carried
out by comparing the different {\bf k}-points assured that the final results
were well converged both in their overall energetics and in the phonon
spectrum (yielding accuracies of order 2~cm$^{-1}$). The 22-atom supercell
has 63 optical modes. However, to illustrate the displacement patterns of
the phonon modes deduced from real-space translational invariance, in
Figs.~1--3 we show our results in the 44-atom $I4/m$ cell.

\begin{table*}[t]
%\begin{ruledtabular}
\caption{Symmetry analysis for space group $I4/m$ and assignment of selected
optical modes in K$_{0.8}$Fe$_{1.6}$Se$_{2}$. The "=", "$\perp$", and "$\angle$"
symbols denote respectively eigenmode directions parallel, perpendicular, and
at an angle to the FeSe plane of the crystal.}
\begin{center}
\begin{tabular*}{15cm}{@{\extracolsep{\fill}} cccccccccc}
\hline \hline
Atom & Wyckoff & \multicolumn{7}{c}{Optical modes} & \\
\cline{3-9}
 & position & \multicolumn{3}{c}{Raman active} & \multicolumn{4}{c}{Infrared
active} & \\
\hline
K  & 8h & \multicolumn{3}{c}{$2A_{g}+2B_{g}+2E_{g}$} & \multicolumn{4}{c}{$A_{u}
 + 4E_{u}$} \\
Fe & 16i & \multicolumn{3}{c}{$3A_{g}+3B_{g}+6E_{g}$} & \multicolumn{4}{c}{$3A_{u}
+ 6E_{u}$} \\
Se & 4e & \multicolumn{3}{c}{$A_{g}+2E_{g}$} & \multicolumn{4}{c}{$A_{u}+2E_{u}$}
\\
Se & 16i & \multicolumn{3}{c}{$3A_{g}+3B_{g}+6E_{g}$} & \multicolumn{4}{c}{$3A_{u}
+ 6E_{u}$} \\
\\
Cal. Freq. & Expt. Freq. & Symmetry & Index & Atoms & \multicolumn{4}{c}
{Direction of eigenmode} \\
\cline{6-9}
(cm$^{-1}$) & (cm$^{-1}$) & & & & K(8h) & Fe(16i) & Se(4e) & Se(16i)\\
66.7 & 61.4 & $B_{g}$ & $^1B_{g}$ & Se &  & & & $\angle$ \\
75.1 & 66.3 & $A_{g}$ & $^1A_{g}$ & Se & & & $\perp$ & \\
106.2 & 100.6 & $B_{g}$ & $^2B_{g}$ & K & $\perp$ & & \\
130.5 & 123.8 & $A_{g}$ & $^2A_{g}$ & Se & & & & $\angle$ \\
159.2 & 134.6 & $A_{g}$ & $^3A_{g}$ & Se & & & $\perp$ & $\angle$ \\
149.0 & 141.7 & $B_{g}$ & $^3B_{g}$ & Se & & & & $\angle$ \\
212.6 & 202.9 & $A_{g}$ & $^4A_{g}$ & Se & & & $\perp$ & $\angle$ \\
238.3 & 214.3 & $B_{g}$ & $^4B_{g}$ & Fe & & $\perp$ & & \\
268.5 & 239.4 & $A_{g}$ & $^5A_{g}$ & Fe & & $\angle$ & & \\
286.1 & 264.6 & $A_{g}$ & $^6A_{g}$ & Fe,Se & & $\angle$ & $\perp$ & \\
279.0 & 274.9 & $B_{g}$ & $^5B_{g}$ & Fe & & $\angle$ & & \\
83.3 & & $E_g$ & & K,Se & $\perp$ & & & $\angle$ \\
102.4 & & $E_g$ & & K,Se & $\perp$ & & = & = \\
143.4 & & $E_g$ & & Se & & & = & = \\
208.7 & & $E_g$ & & Se & & & = & $\angle$ \\
242.5 & & $E_g$ & & Fe,Se & & $\angle$ & = & \\
284.9 & & $E_g$ & & Fe,Se & & $\angle$ & = & \\
119.1 & 102.2$^{a}$ & $A_{u}$ & & K & $\perp$ & & & \\
212.3 & 208.3$^{a}$ & $A_{u}$ & & Se & & & $\perp$ & $\angle$ \\
253.4 & 236.3$^{a}$ & $A_{u}$ & & Fe & & $\angle$ & & \\
308.5 & & $A_{u}$ & & Fe,Se & & $\angle$ & $\perp$ & \\
67.0 & & Chiral & & Se & & & & $\angle$ \\
86.2 & & Chiral & & K & = & & & \\
301.3 & & Chiral & & Fe & & = & & \\
269.6 & & Breathing & & Fe & & $\angle$ & & \\
\hline \hline
$^{a}$Ref.~\onlinecite{wang11}.\\
\end{tabular*}
\end{center}
\end{table*}

Calculated phonon frequencies for prominent modes of all symmetries are
listed in Table I. The experimental frequencies are discussed in Sec.~IV.
As expected, the majority of the modes are vibrations related to the Fe and
Se atoms in the primary structural unit, and this includes all but one of
the experimentally relevant modes (Table I). Vibrations of the K atoms
appear only at low energies, reflecting the weak restoring forces they
encounter far from the FeSe planes. In the real material, these atoms are
thought to be rather mobile.

\begin{table*}[t]
%\begin{ruledtabular}
\caption{Eigenvectors of the most important atoms involved in selected
vibrational modes of K$_{0.8}$Fe$_{1.6}$Se$_{2}$ (Table I). Directions $x$,
$y$, and $z$ are those shown in Figs.~1$-$4.}
\begin{center}
\begin{tabular*}{16cm}{@{\extracolsep{\fill}} cccccccccccccccc}
\hline \hline
Cal. Freq. & Symmetry & \multicolumn{4}{c}{Eigenvector (x y z)} \\
\cline{3-6}
(cm$^{-1}$) & & K(8h) & Fe(16i) & Se(4e) & Se(16i)\\
\hline
66.7 & $B_{g}$ & & & & (-0.07 0.24 0.20) \\
75.1 & $A_{g}$ & & & (0.00 0.00 0.34) & \\
106.2 & $B_{g}$ & (0.00 0.00 0.39) & & \\
130.5 & $A_{g}$ & & & & (0.20 0.12 0.20) \\
159.2 & $A_{g}$ & & & (0.00 0.00 0.29) & (-0.07 0.22 0.11) \\
149.0 & $B_{g}$ & & & & (0.11 0.13 0.29) \\
212.6 & $A_{g}$ & & & (0.00 0.00 0.40) & (-0.04 0.11 0.23) \\
238.3 & $B_{g}$ & & (0.00 0.04 0.31) & & \\
268.5 & $A_{g}$ & & (0.10 0.27 0.12) & & \\
286.1 & $A_{g}$ & & (0.15 0.08 0.23) & (0.00 0.00 0.24) & \\
279.0 & $B_{g}$ & & (-0.15 0.12 0.19) & & \\
83.3  & $E_g$   & (0.00 0.00 0.17) & & & (0.12 0.14 0.05) \\
102.4 & $E_g$   & (0.00 0.00 0.28) & & (-0.12 0.22 0.00) & (-0.03 0.24 0.01) \\
143.4 & $E_g$   & & & (-0.25 0.18 0.00) & (-0.11 0.29 0.04) \\
208.7 & $E_g$   & & & (0.34 -0.04 0.00) & (0.24 -0.04 0.15) \\
242.5 & $E_g$   & & (0.19 -0.03 0.22) & (-0.06 0.26 0.00) & \\
284.9 & $E_g$   & & (0.24 -0.02 0.07) & (-0.19 0.18 0.00) & \\
119.1 & $A_{u}$ & (0.00 0.00 -0.46) & & & \\
212.3 & $A_{u}$ & & & (0.00 0.00 -0.45) & (0.13 0.22 0.08) \\
253.4 & $A_{u}$ & & (0.03 0.27 -0.16) & & \\
308.5 & $A_{u}$ & & (0.01 0.17 0.22) & (0.00 0.00 -0.23) & \\
67.0 &  Chiral  & & & & (0.20 0.23 0.06) \\
86.2 &  Chiral  & (0.02 0.42 0.00) & & & \\
301.3 & Chiral  & & (0.01 0.29 0.00) & & \\
269.6 & Breathing & & (-0.08 0.27 0.09) & & \\
\hline \hline
\end{tabular*}
\end{center}
\end{table*}

The atomic displacement patterns of the assigned $A_g$ and $B_g$ modes are
shown respectively in Figs.~1 and 2. The displacement arrows are to scale
between panels, and it is clear that the largest atomic motions are in
the $c$-direction, while in-plane motion is more restricted. The right-hand
columns of Table I detail the character of the calculated eigenmodes,
showing whether they correspond to atomic motions primarily in the FeSe
plane, perpendicular to it, or in a genuine combination of both. In
addition to the dominant $A_g$ and $B_g$ modes, we also compute a number
of $E_g$ phonons over the same frequency range and list some selected modes
in Table I; these two-fold degenerate modes correspond to in-plane atomic
motions, although they may obtain weak normal components due to the presence
of the vacancies. For reasons of light-scattering selection rules, these
modes are not usually observed in Raman experiments for approximately
tetragonal materials.

The optical phonons we have calculated include not only the Raman-active
modes but also similar numbers of infrared-active ones. Four selected
examples are also listed in Table I and compared to recent experimental
measurements, while their displacement patterns are illustrated in Fig.~3.
As for the Raman-active modes, most of the infrared modes are vibrations
of the Fe and Se atoms. In the tetragonal 122 iron arsenide
superconductors, there exist just two ideal, one-dimensional Raman-active
phonon modes of the FeAs plane, whose symmetries are A$_{1g}$ and
B$_{1g}$,\cite{Iliev} and similarly just two infrared-active modes. To
the extent that K$_{0.8}$Fe$_{1.6}$Se$_{2}$ can be considered as an ordered,
1/5-depleted version of this system, it is clear that the symmetry reduction
and expansion of the unit cell allow many more Raman- and infrared-active
optical modes to exist in the alkali-intercalated FeSe superconductors.

Table II contains the full details of the eigenvectors for the atomic
displacements corresponding to all of our selected modes. While much more
specialized than the polarization information, we provide this data for
completeness and specificity concerning the representations in Figs.~1$-$4.
A full understanding of experimental data concerning resistivity, pairing,
and anomalies in ARPES, inelastic neutron scattering, and magnetic Raman
signals depends on an accurate knowledge of the phonon spectrum, and the
quantitative intensity (Sec.~IV) and polarization information we provide
can be used to calculate the interactions between specific phonon modes
and the itinerant electrons or spin fluctuations of the charge and spin
sectors.

As an immediate example of this, in our calculations we also find some novel
``chiral'' phonon modes, whose atomic displacement patterns are shown in
Fig.~4. These are nondegenerate and primarily in-plane modes in which all
the Fe or Se atoms in a single plane of the structural unit have a net
rotation about the center of the cell, an apparent angular momentum canceled
by the atomic displacements in neighboring unit cells of the same FeSe layer.
The presence of these chiral modes is a direct consequence of chiral
symmetry-breaking in the AFeSe system when the 1/5-depleted vacancy
structure is adopted; the $\sqrt{5}\times\sqrt{5}$ unit cell\cite{wbnd}
itself has an explicit left- or right-handed structural chirality.\cite{Wan11}
The chiral phonon modes are not active in the Raman channel, and therefore are
not observed in the Raman measurements performed here. We suggest, however,
that these chiral modes may be observed by different spectroscopic techniques
in circular polarization configurations.

The presence of chiral modes is of particular interest as a possible probe
of chiral electronic or magnetic excitations, which break time-reversal
symmetry. Such excitations have been discussed in the parent compounds of
cuprate superconductors, where an $A_2$ component attributed to chiral spin
excitations was detected by Raman scattering.\cite{A2} At the theoretical
level, a chiral $d$-density-wave state has also been proposed in the
underdoped state of cuprate superconductors.\cite{DDW} Such modes, involving
chiral electronic motion, may couple preferentially to the chiral phonon
vibration in the 1/5-depleted AFeSe system, in the same way as phonons of
$B_{1g}$ symmetry couple to the $d$-wave superconducting order parameter in
cuprates, and thus Raman phonon spectroscopy may be used to detect their
presence. Without such a coupling to phonon modes, any chiral electronic
or magnetic excitations in the AFeSe system may also be subject to direct
detection by the polarized spectroscopies, such as ARPES and neutron
scattering, also applied in the study of cuprates.

\section{Raman-scattering measurements}

We begin the presentation of our Raman-scattering results by recalling
the basic features the low-temperature spectra, which are shown for
K$_{0.8}$Fe$_{1.6}$Se$_2$ in Fig.~5. The measurements were performed at 9 K
in polarization configurations which separate the $A_g$ and $B_g$ channels.
At least thirteen Raman-active modes are observed, all located below
300 cm$^{-1}$;\cite{ZAM} at least ten infrared-active modes have also
been measured\cite{Infrared} in the same frequency range. This
abundance of optical modes arises due to the symmetry reduction caused
by Fe vacancy ordering, which we have identified as being from D$_{4h}$ to
C$_{4h}$.\cite{ZAM} The space group of the undepleted, 122-type structure,
$I4/mmm$, is reduced to $I4/m$, a process in which all in-plane, two-fold
rotation axes and all mirror planes perpendicular to the ($ab$)-plane are
lost. Both the phonon mode energies and the polarizations observed in
Fig.~5 are in excellent agreement with the calculations of Sec.~III for
the spectrum of optical modes (Table I), as also are the measured infrared
modes.\cite{wang11}

\begin{figure}[t]
\includegraphics[scale=0.4,angle=0]{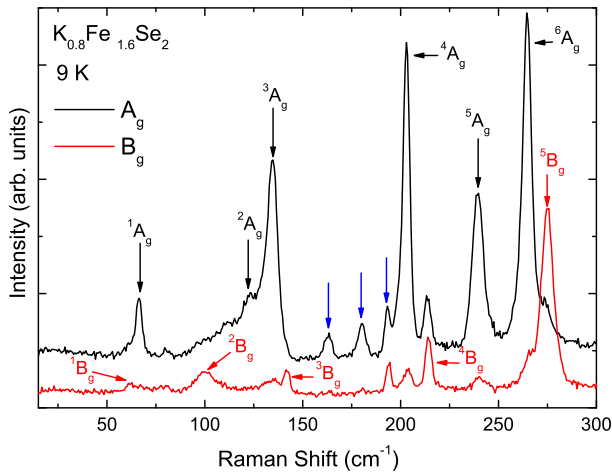}
\caption{(Color online) Raman spectra for K$_{0.8}$Fe$_{1.6}$Se$_2$, measured
in the $A_g$ and $B_g$ channels at 9 K. The mode assignment is made on the
basis of the symmetry analysis and the first-principles calculations of
Sec.~III. The corresponding atomic displacement patterns can be found in
Figs.~1 and 2. Blue arrows indicate unassigned modes.} \label{fig5}
\end{figure}

\subsection{Fe content}

In Fig.~6 we compare the Raman modes in superconducting
K$_{0.8}$Fe$_{1.6}$Se$_{2}$ with those in non-superconducting
KFe$_{1.5}$Se$_{2}$. The modes in the two samples show a general similarity
in intensity and location, which implies a similarity in the microstructures
and symmetries of their FeSe layers. However, it is also evident that
changing the Fe content does cause a significant shift in frequency for most
of the modes. Because these modes are vibrations involving the Fe and Se
ions (Tables I and II), this reflects some significant differences between
the FeSe layers in the two samples. As noted above, neutron diffraction
measurements confirm that the majority phase of the system at an Fe
stoichiometry of 1.6 (20\% Fe vacancies) forms the ideal, four-fold-symmetric,
1/5-depleted, $\sqrt{5}\times\sqrt{5}$ vacancy-ordering pattern. These
measurements also indicate\cite{wbnd} that the same ordering pattern is
maintained for the sample with an Fe content of 1.5, despite the increase
to 25\% Fe vacancies; in this case the 16i Fe positions are only partially
occupied. This occupation means a random distortion of the Fe-Se bonds,
which is responsible for the shifts in mode frequencies. Our results are
thus in agreement with those from neutron diffraction, confirming that
the electronic and magnetic properties of the AFeSe system are rather
sensitive to the vacancy content of the FeSe layers, even if the overall
layer structure is not. These changes should thus be considered as disorder
effects rather than microstructural effects. Altering the Fe content from
1.6 to 1.5 causes our sample to become an insulator with a small gap, which
is estimated by infrared experiments to be 30 meV,\cite{Infrared} and by
transport measurements to be approximately 80 meV.\cite{rblhchgqwl}

\begin{figure}[t]
\includegraphics[scale=0.4,angle=0]{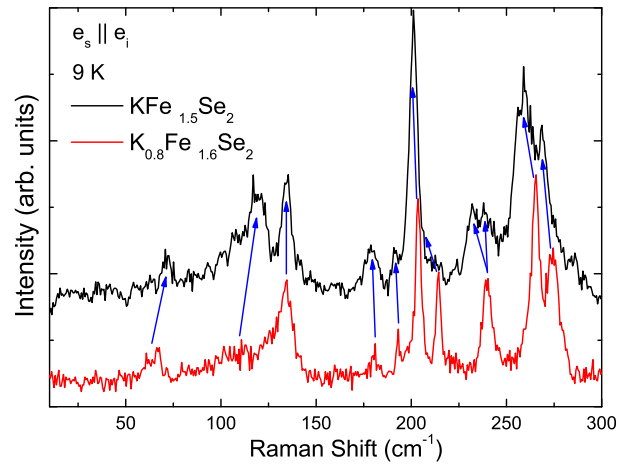}
\caption{(Color online) Comparison between Raman spectra of
K$_{0.8}$Fe$_{1.6}$Se$_2$ and KFe$_{1.5}$Se$_2$. Labels $e_i$ and $e_s$
denote respectively the polarizations of the incident and scattered
light.} \label{fig6}
\end{figure}

\subsection{K substitution}

Raman spectra for the three superconducting crystals K$_{0.8}$Fe$_{1.6}$Se$_2$,
Tl$_{0.5}$K$_{0.3}$Fe$_{1.6}$Se$_2$, and Tl$_{0.5}$Rb$_{0.3}$Fe$_{1.6}$Se$_2$ are
shown in Fig.~7. In contrast to the case of changing Fe content discussed
above, the modes above 60 cm$^{-1}$ exhibit no substantial shift in frequency
(although there are clear differences in relative intensities). This suggests
that substitution within the potassium layers (at fixed Fe content) has little
effect on the FeSe layer, and essentially none on the ordering pattern of the
Fe vacancies. This substitution does, however, cause certain other changes to
occur. The most notable is the presence of some additional phonon modes, which
appear below 60 cm$^{-1}$. These modes can be attributed unambiguously to
vibrations of the heavier Tl and Rb ions, which are absent in the spectra
of K$_{0.8}$Fe$_{1.6}$Se$_2$ (Figs.~5 and 6) and KFe$_{1.5}$Se$_2$ (Figs.~6 and
8). The other important alteration is the dramatic intensity enhancement of
the mode at 180 cm$^{-1}$, which has $A_g$ character but cannot (Fig.~5) be
assigned well from the calculations of Sec.~III; we discuss this feature in
detail below.

\begin{figure}[t]
\includegraphics[width=8.2cm,angle=0]{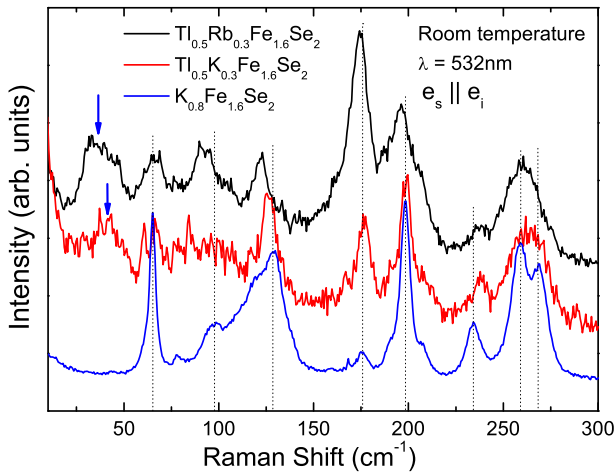}
\caption{(Color online) Raman spectra of the three superconducting crystals
at room temperature. Additional modes, indicated by blue arrows, appear below
60 cm$^{-1}$ in the Tl- and Rb-substituted samples. Dotted lines are guides
indicating the peak positions.}
\label{fig7}
\end{figure}

The additional low-frequency modes induced by K-substitution are shown in
Fig.~8. These become weaker but not narrower with decreasing temperature,
eventually disappearing at 9 K. This behavior is similar to that of the
66 cm$^{-1}$ Se $A_g$ mode, which largely follows the Bose-Einstein thermal
factor.\cite{ZAM} By comparison with K$_{0.8}$Fe$_{1.6}$Se$_2$, these modes may
readily be identified as vibrations of heavier Tl and Rb ions. It should also
be noted here that there exist two possible Wyckoff positions for the A ions,
namely 2a and 8h, and that no Raman-active modes are allowed for atoms in the
2a positions. No structural transition is found below the N\'eel temperature
(520 K) in neutron-diffraction studies of K$_{0.8}$Fe$_{1.6}$Se$_2$,\cite{wbnd}
and the temperature-dependent Raman spectra in Fig.~8 show that it is
reasonable to assume the same behavior in the Tl- and Rb-substituted
crystals. We therefore deduce that the additional modes are allowed due
to changes of the local symmetry in the (Tl,K/Rb)-layer, for which a random
occupation of 2a and 8h sites by A ions is the most likely possibility.

\begin{figure}[t]
\includegraphics[width=8.2cm,angle=0]{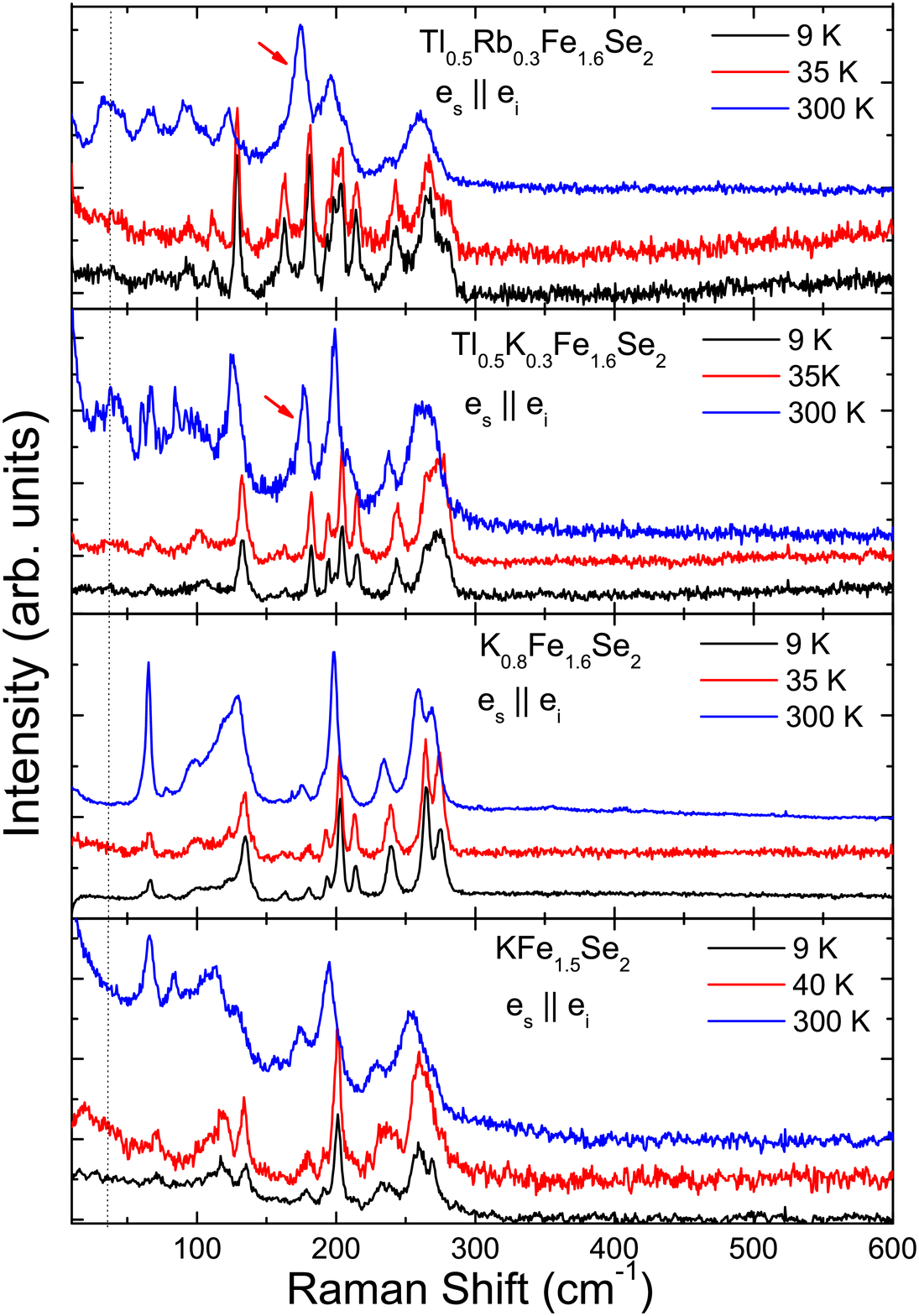}
\caption{(Color online) Raman spectra of superconducting and
non-superconducting AFeSe systems at selected temperatures. The dotted
line indicates the location of additional modes induced by K substitution
and the red arrows indicate modes showing large changes of intensity between
different crystals.}
\label{fig8}
\end{figure}

\subsection{Discussion}

We begin our discussion with the anomalous 180 cm$^{-1}$ mode. This shows not
only a curious temperature-dependence of its intensity between samples, but
also of its frequency at $T_c$. The fact that this mode cannot be assigned
properly by our symmetry analysis and first-principles calculations suggests
that it may be a local mode. One of the most likely candidates for this
would be a nanoscopic region where the Fe vacancy is filled, creating a
locally regular square lattice. Indeed, the As $A_{1g}$ mode in the 122
compounds occurs at a frequency of 182 cm$^{-1}$ in SrFe$_2$As$_2$.\cite{Iliev}

A locally regular square lattice is also one of the leading candidates
suggested in the phase-separation description of the FeSe superconductors.
As noted in Sec.~I, several authors have proposed that the superconducting
minority phase is KFe$_2$As$_2$.\cite{Chen,Li,Friemel} This scenario, that
the 180 cm$^{-1}$ mode we observe is not merely a local filled vacancy, but
the leading fingerprint of a 122 minority phase, would also be consistent
with the jump we observe in the frequency of this mode at $T_c$, which
suggests a strong coupling of this specific mode to the superconducting
order parameter. Further evidence in favor of this interpretation could be
found in the $B_{1g}$ mode of the 122 structure, which appears at 204 cm$^{-1}$
in SrFe$_2$As$_2$.\cite{Iliev} Our results do contain a $B_{g}$ component very
close to this frequency, but we caution that it is accompanied by a very
strong $A_{g}$ signal, and may only be a shadow of this mode arising due to
a disorder-induced mixing of local symmetries.\cite{Kakihana}

\begin{figure}[t]
\includegraphics[width=7.0cm]{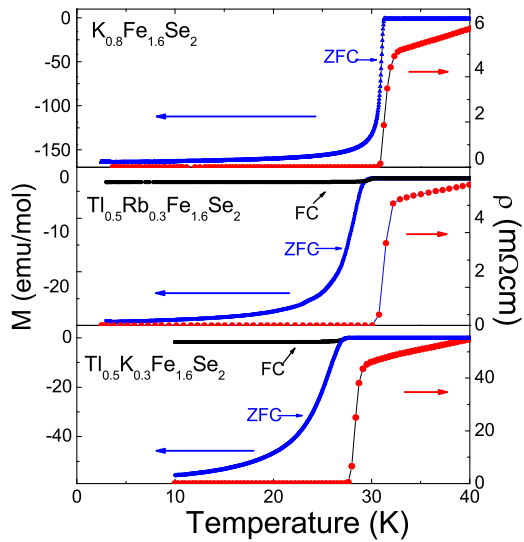}
\caption{(Color online) Superconducting and diamagnetic transitions for the
three superconducting crystals used in the Raman-scattering investigations.}
\label{fig0}
\end{figure}

In the general context of phase separation, it is clear that our samples
have both a robust structural and magnetic order (from the neutron diffraction
studies performed on the same crystals) and a clear superconducting component.
We have been unable to find any evidence for the presence of secondary phases
in X-ray and neutron-scattering studies,\cite{Xray,wbnd} and we show in
Fig.~9 that the resistive and diamagnetic transitions at the onset of
superconductivity are sharp and continuous in all three samples. However,
as pointed out by many authors, none of these results is sufficient to
exclude minority phases with a low volume fraction, and the data for the
superconducting transitions show only that the percolation of the
superconducting fraction is complete and homogeneous (which would be
consistent with a microscale phenomenon).

\begin{figure}[t]
\includegraphics[width=8.2cm,angle=0]{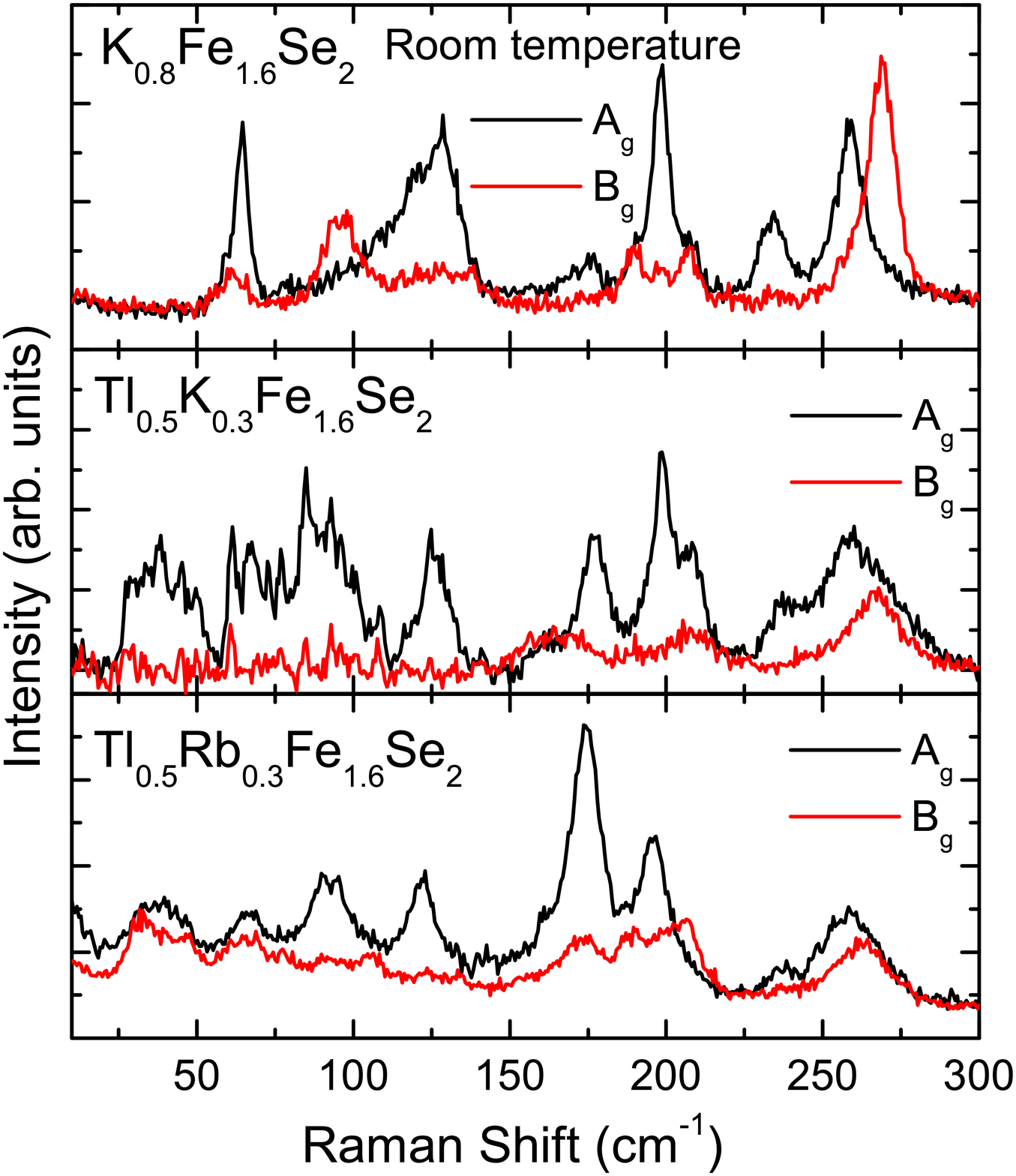}
\caption{(Color online) Polarized Raman spectra of the three superconducting
crystals, showing that phonon widths are generally lower in
K$_{0.8}$Fe$_{1.6}$Se$_{2}$.}
\label{fig9}
\end{figure}

In the most extreme version of a phase-separation scenario, only the 122
and 245 (A$_{0.8}$Fe$_{1.6}$Se$_2$) phases would exist, and altering Fe content
would affect only their ratio. While this situation would account for a loss
of superconductivity on reducing the Fe content, it does seem to require at
least one further low-Fe phase in the doping range of our samples. Our results
do not support this scenario. It would predict that changes in the Fe
stoichiometry in Fig.~6 should appear only as alterations in phonon intensity,
rather than to the phonon frequencies as we observe. Our results definitely
indicate continuous alterations to a single majority phase, and show further
that some vacancy-disorder effects are clearly (if not strongly) detectable.
Thus we conclude that our Raman phonon spectra contain no unambiguous evidence
for a robust 122 minority phase, and we suggest rather a phase-separation
scenario in which the minority phase is one of homogeneous vacancy disorder.

Returning now to the anomalous phonon modes, both scenarios (a secondary 122
phase and locally filled Fe vacancy sites) can account qualitatively for the
frequencies of additional phonon modes beyond our dynamical analysis. While
both also explain the anomalous behavior of the 180 cm$^{-1}$ mode at $T_c$,
neither accounts directly for the anomalous intensity of this mode. To
explain this, we note that the assignment of the mode as a (local or bulk)
version of the 122 $A_{1g}$ mode means that it involves a $c$-axis displacement
of the Se atoms. These are the FeSe modes most strongly affected (Figs.~7 and
8) by A-induced changes in the local microstructure, and we suggest that these
alter the mode intensity in the same way as for the 66 cm$^{-1}$ mode.\cite{ZAM}

Away from the nature of the phase separation, we comment also on the
distinctive low-energy background observed for the four crystals we have
measured (Fig.~8). The spectrum of semiconducting KFe$_{1.5}$Se$_2$ at room
temperature rises strongly at low frequencies, whereas the low-energy part
for K$_{0.8}$Fe$_{1.6}$Se$_2$ is rather flat. All of the background contributions
fall with decreasing temperature. We suggest that the low-frequency
enhancement may be due to electronic Raman scattering. Because
KFe$_{1.5}$Se$_2$ is a small-gap semiconductor, the smaller Coulomb screening
effect relative to a normal metal would allow stronger charge-density
fluctuations and hence a larger electronic Raman scattering contribution.

Finally, one of the most surprising features of the K$_{0.8}$Fe$_{1.6}$Se$_2$
material that makes up the majority of our samples is its apparently high
degree of structural order. This occurs despite its depleted nature, which
one would expect to be prone to atomic disorder. Evidence for disorder
can in fact be found in the polarized Raman spectra at room temperature
(Fig.~10). The Tl- and Rb-substituted samples show larger phonon widths
compared to K$_{0.8}$Fe$_{1.6}$Se$_2$, which implies that more disorder is
induced by the substitution. Given that no substantial shifts occur in the
mode frequencies for the three samples, this disorder can be attributed to
the random occupation and motion of the K, Tl, and Rb ions. The breaking
of local symmetry and periodicity in the A layer acts to shorten the phonon
lifetimes also in the FeSe layer. Overall, it appears that the high stability
of the $\sqrt{5}\times\sqrt{5}$ vacancy-ordered structure, which assures the
constant frequencies of the phonon modes we observe in all our superconducting
samples, may be a consequence of the very specific magnetically ordered state
it allows.

\section{Summary}

To conclude, we have measured Raman spectra in single-crystalline
samples of the superconductors K$_{0.8}$Fe$_{1.6}$Se$_2$,
Tl$_{0.5}$K$_{0.3}$Fe$_{1.6}$Se$_2$, and Tl$_{0.5}$Rb$_{0.3}$Fe$_{1.6}$Se$_2$,
as well as in their insulating derivative compound KFe$_{1.5}$Se$_2$. A
symmetry analysis and first-principles calculations of the zone-center
phonons, both based on the $\sqrt{5}\times \sqrt{5}$ vacancy-ordering
pattern of the K$_{0.8}$Fe$_{1.6}$Se$_2$ unit cell, allow an excellent
assignment of the observed phonon modes. We illustrate the corresponding
atomic displacement patterns and demonstrate the presence of chiral phonon
modes.

We observe a clear frequency shift in all phonons between superconducting
K$_{0.8}$Fe$_{1.6}$Se$_2$ and non-superconducting KFe$_{1.5}$Se$_2$, showing
the effect of further Fe vacancies within the $\sqrt{5}\times \sqrt{5}$
structure on the microscopic properties of the FeSe layers. By contrast,
the frequencies of modes involving Fe and Se ions are little affected on
substituting K by Tl or Rb. However, this substitution does induce additional
Tl and Rb modes below 60 cm$^{-1}$. Our measurements also contain a number of
anomalies, which may be purely effects of the intrinsic vacancy disorder or
may be explained in part by the presence of the weak minority phase
responsible for superconductivity. Our results reveal the complex effects
of Fe vacancies in the FeSe plane, laying the foundation for a full
understanding of the distinctive structural, electronic, magnetic, and
superconducting properties of the A$_{x}$Fe$_{2-y}$Se$_2$ series of materials.

\acknowledgments

We thank W. Bao and Z. Y. Lu for helpful discussions. This work was
supported by the 973 program of the MoST of China under Grant
No.~2011CBA00112, by the NSF of China under Grant Nos.~11034012 and
11004243, by the Fundamental Research Funds for Central Universities, and
by the Research Funds of Renmin University of China (RUC) under grants
10XNI017, 10XNL016, and 08XNF018. Computational facilities were provided
by the HPC Laboratory in the Department of Physics at RUC. The atomic
structures and displacement patterns were plotted using the program
XCRYSDEN.~\cite{kokalj}

\end{document}